\documentclass[twocolumn,pre,amsmath,amssymb,superscriptaddress,longbibarriers,english]{revtex4-1}
\usepackage{array}
\usepackage{tabularx}
\usepackage{bm}
\usepackage{graphics}
\usepackage{amsmath}
\usepackage{amssymb}
\usepackage{epsfig}
\usepackage{graphicx}
\usepackage{float}
\usepackage[usenames,dvipsnames]{color}
\usepackage{mathtools}
\usepackage{marginnote}
\usepackage{etoolbox}
\usepackage{lipsum}
\usepackage{color}
\usepackage{hyperref}
\usepackage{comment}

\usepackage{braket}
\newcommand{\kvec}{\mathbf{k}}
\newcommand{\rvec}{\mathbf{r}}

\UseRawInputEncoding

\usepackage{ae}
\usepackage{aecompl}
\usepackage{mathrsfs}
\usepackage[T1,OT1]{fontenc}

\DeclareTextCommand{\DJ}{OT1}{%
  \raisebox{-0.1ex}{\scalebox{0.75}[1.4]{--}}\kern-.4em D%
}

\newcommand{\br}{{\bf r}}

\begin{document}
\title{Active Self-Consistent Field Theory for Ornstein-Uhlenbeck Polymers}
\author{Yuliang Huang}
\address{School of Physics, Key Laboratory of Quantum Materials and Devices of Ministry of Education, Southeast University, Nanjing 211189, China}
\author{Chun-Lai Ren}
\address{National Laboratory of Solid State Microstructures and School of Physics, Collaborative Innovation Center of Advanced Microstructures, Nanjing University, Nanjing 210093, China}
\author{Qiyun Tang}
\altaffiliation{qtang@seu.edu.cn}
\address{School of Physics, Key Laboratory of Quantum Materials and Devices of Ministry of Education, Southeast University, Nanjing 211189, China}

\begin{abstract}
We develop an active self-consistent field theory (ASCFT) for studying the steady-state behavior of active Ornstein--Uhlenbeck polymers. Starting from the stochastic equations of motion under the unified colored noise approximation, we derive an effective Hamiltonian that extends the classical polymer field theory to non-equilibrium systems. The resulting free energy functional incorporates both the Flory--Huggins interaction parameter $\chi N$ and the persistence time $\tau$ of the active noise, enabling a unified description of thermodynamic and activity-driven effects. To solve the governing equations, we implement a stable implicit-explicit numerical scheme that handles the fourth-order term induced by activity. Our simulations reveal that increasing activity suppresses microphase separation, with the density modulation amplitude decaying as $\Delta\phi \propto \tau^{-1/2}$ for large $\tau$. This scaling is independent of copolymer composition and is confirmed by an asymptotic analysis of the free energy functional in the large-$\tau$ limit. The ASCFT framework provides a new theoretical tool for predicting and designing the non-equilibrium morphologies of active polymer systems, bridging the gap between traditional self-consistent field theory and active matter physics. 
\end{abstract}

\maketitle

\section{Introduction}

Active polymers represent a class of non-equilibrium soft matter systems, where internal energy consumption or environmental driving sustains steady states characterized by persistent active stresses \cite{Sarkar2014,Isele-Holder2015,Isele-Holder2016,Mousavi2019,Shee2020,Eisenstecken2022,Sarkar2023,Liverpool2001,Jayaraman2012,Ghosh2014,Harder2014,Kaiser2014,Shin2015,Eisenstecken2017,Winkler2017,Smrek2017,Philipps2022}. These stresses arise either from self-propelled monomeric units (e.g., Active Brownian Particles, ABPs) \cite{Sarkar2014,Isele-Holder2015,Isele-Holder2016,Mousavi2019,Shee2020,Eisenstecken2022,Sarkar2023} or from immersion in an active bath \cite{Liverpool2001,Jayaraman2012,Ghosh2014,Harder2014,Kaiser2014,Shin2015,Eisenstecken2017,Winkler2017,Smrek2017,Philipps2022}. Coarse-grained models typically describe this activity through non-Markovian stochastic processes, most notably the Active Ornstein-Uhlenbeck Process (AOUP) \cite{Kim1997,Maggi2014,Marconi2015,Fodor2016a,Mandal2017,Bonilla2019,Caprini2019,Sevilla2019,Flenner2020,Woillez2020,
Bothe2021,Martin2021a,Crisanti2023,Dutta2023,Semeraro2023,Singha2023,Rangaig2024,Wittmann2025,Maggi2015,Paoluzzi2020,Tang2024c}, which captures the essential feature of exponentially correlated noise. Activity couples intimately with polymer connectivity, leading to rich dynamical and structural phenomena. For tangentially driven polar chains, activity dramatically enhances long-time diffusion while leaving equilibrium conformations unchanged \cite{Philipps2022}. More generally, active forces modify chain relaxation, swell or shrink conformations depending on stiffness, and induce distinct dynamical regimes from ballistic to enhanced diffusive motion \cite{Mousavi2019,Eisenstecken2017}. Beyond single-chain physics, activity can drive collective phase separation in mixtures, a mechanism with profound implications for understanding biological organization, such as the segregation of active (euchromatin) and passive (heterochromatin) chromatin domains in the cell nucleus \cite{Jost2014,Lieberman-Aiden2009,Solovei2016}. Thus, active polymers bridge fundamental non-equilibrium physics with the dynamics of biological filaments and the design of novel synthetic active materials.

Significant theoretical progress has been made in understanding AOUP-driven polymers \cite{Mousavi2019,Shee2020,Eisenstecken2022,Sarkar2023,Eisenstecken2017}. Analytical solutions of linear Langevin equations for Gaussian semiflexible chains reveal how colored active noise renormalizes relaxation times, effective stiffness, and fluctuations \cite{Mousavi2019,Eisenstecken2017}. Formal mappings between AOUP/ABP trajectories and equilibrium polymer configurations provide powerful analogies for computing moments \cite{Shee2020}, and path-integral formulations have been developed to describe stationary-state distributions \cite{Eisenstecken2022}. Studies of AOUP-driven harmonic chains further show that bulk steady-state properties can become universal, independent of the specific active noise details \cite{Sarkar2023}. Crucially, in mixtures, activity-driven phase separation exhibits a polymer-length-dependent critical point reminiscent of equilibrium Flory-Huggins theory, where the critical activity scales inversely with chain length \cite{Smrek2017}. This finding strongly suggests the need for a statistical field theory that can treat the competition between active driving and polymer entropy-enthalpy balance.

In passive polymer systems, self-consistent field theory (SCFT) has long served as the cornerstone theoretical framework for predicting equilibrium morphologies in polymer systems, especially block copolymers \cite{Edwards1965,Bates1990,Matsen1994,Fredrickson1996,Schmid1998,Drolet1999,Tang2009,Tang2010,Glaser2014,Li2014a,
Beardsley2016,Sun2017,Vorselaars2020,Matsen2023,Fredrickson2023,Geng2025}. By balancing entropic chain stretching with enthalpic interactions, SCFT successfully captures rich phase behavior from lamellar and cylindrical structures to gyroid networks \cite{Bates1990,Matsen1994,Fredrickson1996}. However, a self-consistent field theory framework that systematically incorporates the finite persistence time of AOUP noise into a polymer free energy functional is entirely lacking. Existing particle-based models and analytical treatments \cite{Mousavi2019,Shee2020,Eisenstecken2022,Sarkar2023} are not formulated to predict the phase behavior of active polymeric systems from a unified field-theoretic perspective. This gap motivates the development of a mean-field platform to explore how active persistence and thermodynamic interactions jointly sculpt non-equilibrium steady-state structures.

In this work, we bridge this gap by developing an Active Self-Consistent Field Theory (ASCFT) for Ornstein-Uhlenbeck polymers. Our approach begins with the stochastic differential equations governing active OU particles and employs the Unified Colored Noise Approximation (UCNA) \cite{Maggi2015,Paoluzzi2020} to derive a steady-state probability distribution. From this distribution, we construct an effective Hamiltonian that explicitly includes both passive interactions and active contributions, thereby extending the SCFT formalism to non-equilibrium steady states. The resulting effective free energy functional incorporates the persistence time $\tau$ of the active noise as a key parameter, alongside the conventional Flory-Huggins parameter $\chi N$.

To solve the modified diffusion equations arising from the active free energy functional, we devise a stable and efficient numerical algorithm based on an implicit-explicit (IMEX) operator splitting scheme. This method handles the negative diffusion coefficient in the fourth-order term (induced by activity) implicitly, ensuring numerical stability while maintaining computational tractability. Using this algorithm, we compute steady-state density profiles for both symmetric and asymmetric block copolymers, and our results demonstrate that increasing activity $\tau$ suppresses the amplitude of density modulations, leading to a homogenization of the system. The density difference $\Delta\phi_A$ follows a power-law decay $\Delta\phi_A \propto \tau^{-1/2}$ for $\tau \gtrsim 1$, a scaling behavior that is independent of the block fraction $f$ at large $\tau$. We complement these numerical findings with an asymptotic analysis of the free energy functional in the large-$\tau$ limit, deriving the $\tau^{-1/2}$ scaling analytically and confirming the universality of the exponent.

By integrating stochastic active dynamics into a mean-field polymer theory, ASCFT provides a new paradigm for studying non-equilibrium polymer systems. It not only captures the suppression of phase separation by activity but also predicts the emergence of transient structures and scaling laws that are inaccessible to passive SCFT. This work lays the foundation for future studies on more complex active polymer systems, including those with hydrodynamic interactions, multi-component mixtures, and time-dependent activity patterns, offering a versatile tool for designing and controlling active soft materials.

\section{Method Details}\label{Method}
Here we first derive the probability distribution of steady state for active Ornstein-Uhlenbeck particles from the stochastic process and the Fokker-Planck equation. Subsequently, we construct the effective Hamiltonian from the steady state probability, and extend this effective Hamiltonian to the self-consistent field theory for active polymers. 

\subsection{Steady State Probability of Active Ornstein-Uhlenbeck Polymers}
Let us consider a set of stochastic differential equations for $n$ active polymers, with each chain containing $N$ monomers: \cite{Maggi2015,Paoluzzi2020}:
\begin{equation}
    \dot{x}_{pq} = -\partial_{x_{pq}}\Phi + \eta_{pq}
    \label{e0}
\end{equation}
where $p=1,\dots,n$ and $q=1,\dots,N$. In the following, we simplify the suffix as $i=p*N+q$, and the stochastic differential equations can be written as:
\begin{equation}
    \dot{x}_{i} = -\partial_{x_{i}}\Phi + \eta_{i}
    \label{e1}
\end{equation}
where \(\partial_{x_{i}} = \partial/\partial x_{i}\) and $\Phi$ is the potential energy experienced by monomer $i$. For polymers, the potential energy includes two contributions: $\Phi(x_i)=U_0(x_i)+U_b(x_i)$, where $U_0(x_i)$ represents the nonbonded contributions, and $U_b(x_i)$ characterizes the bonded potential energies between monomer $i$ and its neighboring bonded monomers $i+1$ and $i-1$.

For active Ornstein-Uhlenbeck polymers, \(\eta_{i}\) characterizes a set of independent Gaussian colored-noises with exponential time-correlation and zero mean values:
\begin{equation}
\begin{gathered}
    \langle\eta_{i}(t)\eta_{j}(s)\rangle = \delta_{ij}\frac{D}{\tau}e^{-|t-s|/\tau}\\
    \langle\eta_{j}\rangle=0
\end{gathered}
\end{equation}
which can be produced by the Ornstein-Uhlenbeck process:
\begin{equation}
    \tau\dot{\eta}_{i} = -\eta_{i} + \sqrt{D}\Gamma_{i}    
    \label{e2}
\end{equation}
here $\tau$ is the characteristic time for the exponential time-correlation, which depicts the persistence of activity induced by external input. $D$ is the self-diffusion coefficient of active particles. \(\Gamma_{i}\) characterizes a set of white-noises:
\begin{equation}
\begin{gathered}
    \langle\Gamma_{i}(s)\Gamma_{j}(t)\rangle =2 \delta_{ij}\delta(t-s)\\
    \langle\Gamma_{i}\rangle=0
\end{gathered}
\end{equation}

Before proceeding further, we nondimensionalize the system using a characteristic length \(l_0\) and the relaxation time \(\tau_R = l_0^2 / D\). Introduce the dimensionless variables:
\begin{equation}
\tilde{x} = \frac{x}{l_0},\quad \tilde{t} = \frac{t}{\tau_R},\quad \tilde{\Phi} = \frac{\Phi}{k_B T},\quad \tilde{\eta} = \frac{\eta}{l_0/\tau_R},\quad \tilde{\tau} = \frac{\tau}{\tau_R}.
\end{equation}
with this choice, \(\tilde{D}=1\) automatically, and the dimensionless versions of \autoref{e1} and \autoref{e2} take exactly the same form as their dimensional counterparts. In the following derivations, we continue to use \autoref{e1} and \autoref{e2} with the understanding that all parameters and variables are dimensionless.

By combining \autoref{e1} and \autoref{e2}, we obtain:
\begin{equation}
    \ddot{x}_{i} + \frac{\partial_{x_{i}}\Phi}{\tau} + \sum_{k}\left(\frac{\delta_{ik}}{\tau} + \partial_{x_{i}x_{k}}\Phi\right)\dot{x}_{k} = \frac{D^{1/2}}{\tau}\Gamma_{i}\label{e5}
\end{equation}

At this stage, two established routes exist to obtain a stationary distribution for active Ornstein-Uhlenbeck particles. 
The first route is the \emph{Unified Colored Noise Approximation} (UCNA) \cite{Maggi2015,Paoluzzi2020}, which neglects the inertial term $\ddot{x}_i$ in \autoref{e5}. The second route retains the inertial term but restricts the potential to a quadratic form (harmonic interactions), where the stationary state can be obtained exactly and satisfies detailed balance \cite{Bonilla2019,Martin2021a}. For non-quadratic potentials, the stationary state exists but is not time-reversal invariant (a non-equilibrium steady state) \cite{Bonilla2019}. Our system is a polymer melt, where monomer motion is strongly overdamped and inertia is negligible on the relevant conformational time scales. Consequently, the UCNA is the physically appropriate and computationally tractable approximation for our system. We therefore adopt the UCNA in the following derivation, which yields:
\begin{equation}
    \dot{x}_{i} = -\sum_{k}[M^{-1}]_{ik}\frac{\partial_{x_{k}}\Phi}{\tau} + \frac{D^{1/2}}{\tau}\sum_{k}[M^{-1}]_{ik}\Gamma_{k}\label{e6}
\end{equation}
Here the symmetric matrix $M$ is defined as \(M_{ij} = \delta_{ij}/\tau + \partial_{x_{i}x_{j}}\Phi\) (\(\partial_{x_{i}x_{k}}\Phi\) represents the Hessian matrix) with its inverse \([M^{-1}]_{ij}\). This is a typical Stratonovich Langevin equation:
\begin{equation}
\dot{x}_{i} = A_i(x) + \sum_{j}B_{ij}(x)\Gamma_{j}
\end{equation}

The corresponding Fokker-Planck equation is:
\begin{equation}
    \dot{\Omega} =-\sum_i \partial_{x_{i}} (A_i\Omega) + \frac{1}{2}\sum_{i,j,k}\partial_{x_{i}}\partial_{x_{j}}(B_{ik}B_{jk}\Omega)
\end{equation}

By replacing $A_i$ and $B_{ij}$ with the associated terms in \autoref{e6}, we obtain the Fokker-Planck equation as:
\begin{equation}
\begin{split}
    \dot{\Omega} = &\sum_{l,i}\partial_{x_{i}}\left\{\frac{1}{\tau}[M^{-1}]_{li}\left(\partial_{x_{i}}\Phi\right)\Omega\right\} \\
    &+ \frac{D}{\tau^{2}}\sum_{l,i,j}\partial_{x_{i}}\left\{[M^{-1}]_{li}\,\partial_{x_{j}}\left([M^{-1}]_{ij}\,\Omega\right)\right\}\label{e7}
\end{split}
\end{equation}
where \(\Omega = \Omega(x_{1},\dots,x_{N})\) is the probability density function. 

The zero-flux stationary solution of \autoref{e7}, $\Omega=P_{ss}$, is obtained by setting:
\begin{equation}
\begin{split}
    &\sum_{i}[M^{-1}]_{li}\left(\partial_{x_{i}}\Phi\right)P_{ss} \\
    &+ \frac{D}{\tau}\sum_{i,j}[M^{-1}]_{li}\,\partial_{x_{j}}\left\{[M^{-1}]_{ji}P_{ss}\right\} = 0
\end{split}
\end{equation}

By multiplying with \(M_{kl}\) and summing over \(l\) we get:
\begin{equation}
    -\frac{\tau}{D}(\partial_{x_{k}}\Phi)P_{ss} - P_{ss}\sum_{j}\partial_{x_{j}}[M^{-1}]_{jk} = \sum_{j}[M^{-1}]_{jk}\,\partial_{x_{j}}P_{ss}\label{e9}
\end{equation}
where we have used the identity \(\sum_{l}M_{kl}[M^{-1}]_{li} = \delta_{ki}\). Repeating this step by multiplying with \(M_{nk}\) and summing over $k$ both sides of \autoref{e9}:

\begin{equation}
    -P_{ss}\frac{\tau}{D}\sum_{k}M_{nk}(\partial_{x_{k}}\Phi) + P_{ss}\sum_{jk}[M^{-1}]_{jk}(\partial_{x_{j}}M_{kn}) = \partial_{x_{n}}P_{ss}\label{e10}
\end{equation}
where we have incorporated the following express into the second term on the l.h.s of \autoref{e9}
\begin{equation}
\begin{gathered}
\partial_{x_{j}}\left(\sum_{k}[M^{-1}]_{jk}M_{kn}\right) = \partial_{x_{j}}(\delta_{jn}) = 0\\
\sum_{k}[M^{-1}]_{jk}(\partial_{x_{j}}M_{kn}) = -\sum_{k}(\partial_{x_{j}}[M^{-1}]_{jk})M_{kn}
\end{gathered}
\end{equation}

Because \(M_{ij}\) contains the \(2^{\text{nd}}\)-order derivatives of the potential, one has \(\partial_{x_{j}}M_{kn} = \partial_{x_{n}}M_{kj}\), which transform the second term on l.h.s of \autoref{e10} into: 

\[
\sum_{jk}[M^{-1}]_{jk}(\partial_{x_{j}}M_{kn}) = \sum_{jk}[M^{-1}]_{jk}(\partial_{x_{n}}M_{kj}) = \frac{1}{|M|}\partial_{x_{n}}|M|
\]
where in the last equality \(|M|\) is the determinant of the matrix \(M\) and we have used the Jacobi's formula \((1/|M|)\partial_{x}|M| = \text{Tr}\left(M^{-1}\partial_{x}M\right)\). Returning to \autoref{e9} generates:
\begin{equation}
    P_{ss}\left(-\frac{\tau}{D}\sum_{k}M_{nk}(\partial_{x_{k}}\Phi) + \frac{1}{|M|}\partial_{x_{n}}|M|\right) = \partial_{x_{n}}P_{ss}
\end{equation}
which can be integrated as:
\begin{equation}
    P_{ss} = C\exp\left[-\frac{\Phi}{D} - \frac{\tau}{2D}\sum_{i}(\partial_{x_{i}}\Phi)^{2}\right]||\tau M||
\end{equation}
here $C$ is a normalization factor, and \(||M||\) is the absolute value of the determinant of \(M\). This is the probability distribution of steady state for active Ornstein-Uhlenbeck polymers with the Unified Colored Noise Approximation (UCNA).

\subsection{Constructing the Effective Hamiltonian for Active Polymers}
The steady-state distribution $P_{ss}$ of the active Ornstein-Uhlenbeck polymers yields an effective Hamiltonian, which can be expressed as:
\begin{equation}
\begin{split}
H_{\mathrm{AP}}&=-D\ln P_{\mathrm{ss}}\left(\left\{\mathbf{x}_i\right\}\right) \\
&= \sum_{i}\Phi(x_{i})+\frac{\tau}{2}\sum_{i}(\partial_{x_{i}}\Phi)^{2}-D\sum_{i}\ln\det \tau\mathbf{M}
\end{split}\label{Hamil}
\end{equation}
here the effective Hamiltonian comprises three parts:

\begin{enumerate}
    \item \textbf{Mechanical Interaction ($H_0$)}
    \\
    This includes the nonbonded and bonded interactions
     \begin{equation}
     \begin{split}
    H_0 &= \sum_{i}\Phi(x_{i})= \sum_i [U_0({x_i})+U_b(x_i)]\\
    &=\rho_0\int d \br U_0(\br)+\frac{3n}{2Nb^{2}}\int_{0}^{N}ds\mid\frac{d\br_{\alpha}(s)}{ds}\mid^{2}.
    \end{split}
    \end{equation}
    where we have converted the ``sum over particles'' into a ``spatial integral'' form in the field-theoretic formulation, and the bonded interactions are also converted into continuous description. Here $b$ is the Kuhn length that sets the statistical step of the polymer chain.
    
    \item \textbf{Active Contribution ($H_1$)}
    \\The persistence effects of the self-propulsion introduce an additional energy term. In the original expression, this is a sum over all particles:
     \begin{equation}
          \begin{split}
    H_1 &= \frac{\tau}{2} \sum_i |\nabla_{x_i} \Phi|^2\\
    &=\frac{\tau}{2} \sum_i |\nabla_{x_i} U_0(x_i)+ \nabla_{x_i} U_b(x_i)|^2
        \end{split}
    \end{equation}
    
For Gaussian chains with harmonic bonded potential:
\[
U_b = \frac{3}{2b^2} \left| \frac{d\mathbf{r}}{ds} \right|^2, \quad \nabla U_b = -\frac{3}{b^2} \frac{d^2\mathbf{r}}{ds^2}
\]
Thus, the active contribution \( H_1 \) becomes:

     \begin{equation}
          \begin{split}
H_1 = &\frac{\tau\rho_0}{2} \int d\br \left| \nabla U_0(\br)\right|^2-\frac{3\tau}{b^2}\sum_i\nabla_{x_i} U_0(x_i)\frac{d^2\br_{\alpha}(s)}{ds^2} \\
&+\frac{9n\tau}{2N^2b^4} \int_{0}^{N}ds \left| \frac{d^2\br_{\alpha}(s)}{ds^2} \right|^2 
     \end{split}
    \end{equation}

   \item \textbf{Active Contribution ($H_2$)} 
   \\This term includes a determinant correction:
     \begin{equation}
    H_2 = -D \sum_i \ln \det \tau\mathbf{M}, \quad \tau M_{ij} = \delta_{ij} + \tau \frac{\partial^2 \Phi}{\partial x_i \partial x_j}.
    \end{equation}
    In 2D space, M is a $N\times N$ matrix. To address computational complexity, we employ a small-$\tau$ expansion for the determinant:
    \begin{equation}
    \det\left(\delta_{i j}^{\alpha \gamma}+\tau D_{i j}^{\alpha \gamma}\right) \approx 1+\tau \operatorname{Tr}(D)+O\left(\tau^2\right),
    \end{equation}
    
    where $D_{i j}^{\alpha \gamma}$ is the Hessian matrix of $\Phi$. For ease of conversion into the field formulation, we write it as a spatial integral:
         \begin{equation}
         \begin{split}
         H_2&=-\sum_iD\ln\{||I+\tau H||\}\\
         &=-D\rho_0\int d\br\ln\left[1+\tau\nabla^2\Phi(\br)\right]
         \end{split}
         \end{equation}
    where the logarithmic form is retained, and $D$ is the self-diffusion coefficient of monomers.

    In the small $\tau$, the above $H_2$ can be simplied as:
            \begin{equation}
         H_2=-D\tau\rho_0\int d \br \nabla^2\Phi(\br)
         \end{equation}
This expression can also be obtained from the pertubation method of the steady state disbtribution of active OU particles without the UCNA approximation \cite{Fodor2016a,Martin2021a}. 

Here we note that the potential $\Phi(\br)$ includes two contributions: $\Phi(\br)=U_b(\br)+U_0(\br)$, for the bonded potential, if we choose a harmonic potential $U_b(\br)=(k/2)(\br-\br_0)^2$, which is a good approximation for Gaussian chains, $\nabla^2U_b(\br)=k$ becomes a constant. In this case, the $H_2$ only depends on non-bonded terms. Therefore   
        \begin{equation}
         H_2=-D\tau\rho_0\int d \br \nabla^2U_0(\br)
         \end{equation}
\end{enumerate}

\subsection{Derivation of Self-Consistent Field Theory for Active Polymers}

The local density field for multicomponent systems (e.g., A/B copolymers) \cite{Schmid1998}: 
    \begin{align}
    \hat{\psi}_{A} (\br )&=\frac{1}{\rho_0}\sum_{\alpha=1}^{n} \int_{0}^{fN} d s \, \delta\big( \br-\br_{\alpha} ( s ) \big), \\
    \hat{\psi}_{B} ( \br )&=\frac{1}{\rho_0}\sum_{\alpha=1}^{n} \int_{fN}^{N} d s \, \delta\big( \br-\br_{\alpha} ( s ) \big), 
    \end{align}
    where $f$ is the fraction of A monomers with $\rho_0$ being the average monomer density, and the incompressibility condition requires:
    \begin{equation}
    \hat{\psi}_{A} ( \br )+\hat{\psi}_{B} ( \br )=1, 
    \end{equation}

For a system containing $n$ active OU polymers, the "partition function" can be written as:
\begin{equation}
\begin{split}
Z=\frac{1}{n!} \prod_{\alpha=1}^{n} \int &\mathcal{D} \br_{\alpha} ( s )P[\mathbf{r}_{\alpha} ( s )]\delta [1-\hat{\psi}_A(\mathbf{r})-\hat{\psi}_B(\mathbf{r})] \\&\exp \left[-U_{nb}+H_1+H_2 \right],
\end{split}
\end{equation}
where the conformation $\mathbf{r}_{\alpha}(s)$ of each chain (with $s\in[0,N]$ as the contour parameter) follows Gaussian chain statistics, 
$$
P[\mathbf{r}_{\alpha} ( s )]=\left(\frac{3}{2\pi Nb^{2}}\right)^{3/2}\exp\left[-\frac{3}{2Nb^{2}}\int_{0}^{N}ds\left|\frac{d\mathbf{r}_{\alpha}(s)}{ds}\right|^{2}\right],
$$
which originates from the bonded contributions $U_b$. The non-bonded energy term 
\begin{equation}
U_{nb} =\int d\mathbf{r} \rho_0 U_0(\mathbf{r})=\rho_0\int d \br \chi \hat{\psi}_A(\mathbf{r})\hat{\psi}_B(\mathbf{r})   
\end{equation}
$U_0(\mathbf{r})=\chi \hat{\psi}_A(\mathbf{r})\hat{\psi}_B(\mathbf{r})$ characterizes the ``external potential'' experienced by the monomers at position $\mathbf{r}$. 

To convert the density operator (defined via particle coordinates) into a continuous field, we insert the functional integral representation into the partition function:
\begin{equation}
\begin{split}
    1&=\int\mathcal{D}\psi_A(\br)\delta{\left[\psi_A(\br)-\hat{\psi}_A(\br)\right]}\\&=\int\mathcal{D}\psi_A(\br)\int_{-i\infty}^{i\infty}\mathcal{D}\omega_A(\br)\\&\exp{\left\{\frac{\rho_0}{N}\int d\br\omega_A(\br){\left[\psi_A(\br)-\hat{\psi}_A(\br)\right]}\right\}},
   \end{split}
\end{equation}
    and similarly for $\psi_B(r)$ . The incompressibility condition is expressed as:
\begin{equation}
\begin{split}
\delta&\left[1-\hat{\psi}_A(\br)-\hat{\psi}_B(\br)\right]=\int_{-i\infty}^{i\infty}\mathcal{D}\xi(\br)\\&\exp\left\{\frac{\rho_0}{N}\int d\br\xi(\br)\left[1-\hat{\psi}_A(\br)-\hat{\psi}_B(\br)\right]\right\}.
   \end{split}
\end{equation}
    Here, $\omega_A(r)$, $\omega_B(r)$, and $\xi(r)$ are auxiliary fields that take optimal values during the saddle-point approximation.
    
The partition function can be written as:
 
\begin{equation}
\begin{split}
    Z = &\frac{1}{n!} \int \mathcal{D}\psi_A \mathcal{D}\psi_B \int \mathcal{D}\omega_A \mathcal{D}\omega_B \mathcal{D}\xi \\&\exp \Bigg\{\frac{\rho_0}{N}\int d\br \left[ \xi \left( 1 - \psi_A - \psi_B \right) - \omega_A \psi_A - \omega_B \psi_B \right]\\ &- \frac{\rho_0}{N}\int d\br \, \chi N\psi_A \psi_B + F_{\text{active}} \Bigg\} \prod_{\alpha=1}^n Q[\omega_\alpha, \omega_\beta]
\end{split}
\end{equation}
and the active corrections are explicitly written as
\begin{equation}
\begin{split}
F_{\mathrm{active}}=&\frac{\tau \chi^2 \rho_0}{2}\int d\br \left|\nabla\left(\psi_A(\mathbf{r})\psi_B(\mathbf{r})\right)\right|^2\\&-D\tau \chi \rho_0\int d\br\nabla^2\left[\psi_A(\mathbf{r})\psi_B(\mathbf{r})\right]
\end{split}
\end{equation}

The free energy functional becomes:
\begin{equation}
\begin{split}
    F_{\rm AP} = &-\ln (\frac{Q}{V}) + \frac{1}{V} \int d\br \left\{ \frac{\tau\chi^2 N}{2} \left|\nabla(\psi_A\psi_B) \right|^2\right. \\& - D\tau \chi N\nabla^2(\psi_A\psi_B) +\chi N\psi_A\psi_B \\
    & \left. - \omega_A\psi_A - \omega_B\psi_B  - \xi[1-\psi_A-\psi_B] \right\}
\end{split}\label{FEF}
\end{equation}
This final expression integrates the statistical properties of the Gauss chain model of polymers, the nonequilibrium dynamics of active particles, and the geometric constraints, forming a comprehensive free energy expression for active OU polymers.

\subsection{Self-consistent field equations for Active OU polymers}

The single-chain partition function under external fields is defined as:
\begin{equation}
\begin{split}
    Q&[\omega_A, \omega_B] = \int{\cal D} \br_{\alpha} ( s )P_{\rm eff}[\br_{\alpha} ( s )] \\&\exp \left\{ - \int_0^{f} ds  \omega_A(\br_{\alpha} ( s )) - \int_{f}^{1} ds  \omega_B(\br_{\alpha} ( s )) \right\},
    \label{eq:complexEquation}
    \end{split}
\end{equation}
where the effective single chain weight is:
\begin{equation}
\begin{split}
P_{\text{eff}}[\br_{\alpha}(s)] \propto & \exp\left[ -\int_0^N ds \left(\frac{3}{2Nb^2} \left| \frac{d\br_{\alpha}(s)}{ds} \right|^2 \right.\right.\\&\left.\left. + \frac{9\tau}{2N^2b^4} \left| \frac{d^2\br_{\alpha}(s)}{ds^2} \right|^2 \right)+H_{1,\rm cross} \right]
    \end{split}
\end{equation}
here there is a cross term $H_{1,\rm cross}=-(3\tau/b^2)\sum_i\nabla_{x_i} U_0(x_i)[d^2\br_{\alpha}(s)/ds^2]$ that corrects the single chain weight from both the chain curvature $d^2\br_{\alpha}(s)/ds^2$ and the local non-bonded interactions $\nabla_{x_i} U_0(x_i)$. The later term holds non-zero values only within interfacial region, and these values may be positive or negative, which can be omitted via the statistical average. In this sense, the correction of activity to the single chain weight comes chiefly from the $| (d^2\br_{\alpha}(s)/ds^2)|^2$, which changes the Gaussian chain to a worm-like chain \cite{Shee2020}.    

The single-chain partition function $Q$ is expressed using propagators of the end-segment distribution function  $q(\br,s)$, which satisfies a modified diffusion equation:
    \begin{equation}
\begin{split}
    \frac{\partial q(\br,s)}{\partial s} =& \frac{Nb^2}{6}\nabla^2 q(\br,s)-\frac{N^2b^4\tau}{18}\nabla^4 q(\br,s)\\& - \omega_\alpha(\br) q(\br,s), \quad q(\br,0) = 1,
    \end{split}\label{DF}
    \end{equation}
    where the external fields are:
    \begin{equation}
    \omega_\alpha(\br)=\begin{cases}\omega_A(\br),&0<s<f,\\\omega_B(\br),&f<s<1.\end{cases}
    \end{equation}
    
The modified diffusion \autoref{DF} does not contain the self-diffusion constant \(D\) that appears in the original Langevin equations. This is because \(D\) enters only through the term \(H_2 = -D\sum_i \ln\det(\tau\mathbf{M})\) in the effective Hamiltonian \ref{Hamil}. In the free energy functional \ref{FEF}, this term contributes the spatially integrated expression \(F_2 = -D\tau\chi N\int d\mathbf{r}\,\nabla^2(\psi_A\psi_B)\), which vanishes for periodic boundary conditions. Consequently, \(D\) does not enter either the propagator equation \ref{DF} or the saddle-point equations \ref{39} and \ref{40}.
    
    The adjoint propagator $q^\dagger(\br,s)$ satisfies a similar modified diffusion equation:
     \begin{equation}
\begin{split}
    \frac{\partial q^\dagger(\br,s)}{\partial s} =& -\frac{Nb^2}{6}\nabla^2 q^\dagger(\br,s) +\frac{N^2b^4\tau}{18}\nabla^4 q(\br,s)\\&+ \omega_\alpha(\br) q^\dagger(\br,s), \quad q^\dagger(\br,1)=1    
    \end{split}
    \end{equation}
and the single chain partition function is:
    \begin{equation}
    Q=\int d\br q(\br,1)q^\dagger(\br,1)
    \end{equation}

Applying the saddle-point approximation to the free energy functional yields the self-consistent field equations. For $\omega_A(r)$, we have $\delta F/\delta\omega_A(\br)=0$, which generates:
    \begin{equation}
    \psi_A(\br)=-\frac{V}{Q}\frac{\delta Q}{\delta\omega_A(\br)}=\frac{V}{Q}\int_0^fdsq(\br,s)q^\dagger(\br,s)\label{37}
    \end{equation}
    and similarly for $\omega_B(r)$
    \begin{equation}
    \psi_B(\br)=-\frac{V}{Q}\frac{\delta Q}{\delta\omega_B(\br)}=\frac{V}{Q}\int_f^1dsq(\br,s)q^\dagger(\br,s)\label{38}
    \end{equation}

    These equations determine the density distribution under external fields.

    For $\psi_A(r)$, 
    \begin{equation}
    \begin{split}
    \frac{\delta F}{\delta\psi_A(r)}&=\frac{\partial f}{\partial\psi_A}+[ \frac{\partial f}{\partial u}-\\
    & \nabla\cdot\frac{\partial f}{\partial (\nabla u)}+\nabla^2\frac{\partial f}{\partial (\nabla^2 u)}]\frac{\partial u}{\partial\psi_A}=0
    \end{split}
    \end{equation}
where $u=\psi_{A}(r)\psi_{B}(r)$ and $f=\chi Nu-\omega_A\psi_A-\omega_B\psi_B-\xi[1-\psi_A-\psi_B]+(\tau/2N)\left|\chi N\nabla u\right|^2-D\tau\chi N\nabla^2 u$. This yields the following field equations for $A$ component:
\begin{equation}
\begin{split}
\omega_A(\mathbf{r})=&\chi N\psi_B(\mathbf{r})+\xi(\mathbf{r})\\&-
\tau\chi^{2}N\psi_{B}(\br)\nabla^2\left[\psi_{A}(\br)\psi_{B}(\br)\right]
\end{split}\label{39}
\end{equation}
Here, we note that the $F_2=-D\tau\chi N\nabla^2 u$ term disappears during the variation of free energy functional with respect to $\psi_A(\br)$. Similarly, for $\psi_B(r)$ 
    \begin{equation}
\begin{split}
\omega_B(\mathbf{r})=&\chi N\psi_A(\mathbf{r})+\xi(\mathbf{r})\\&-
\tau\chi^{2}N\psi_{A}(\br)\nabla^2\left[\psi_{A}(\br)\psi_{B}(\br)\right]
\end{split}\label{40}
\end{equation}

    Variation of \autoref{FEF} with respect to $\xi(r)$ gives:
        \begin{equation}
        \frac{\delta F}{\delta\xi(r)}=0\quad\Rightarrow\quad1-\psi_A(r)-\psi_B(r)=0.\label{41}
        \end{equation}

Here the \autoref{37}-\autoref{41} are a set of self-consistent field equations for active OU polymers.

\section{Results and Discussion}

\subsection{Numerical Implementation}

The modified diffusion equation for the forward propagator \( q(\rvec, s) \) in active Ornstein--Uhlenbeck polymer systems reads:
\begin{equation}
    \frac{\partial q(\rvec, s)}{\partial s} = \alpha \nabla^2 q(\rvec, s) + \beta \nabla^4 q(\rvec, s) - \omega(\rvec) q(\rvec, s),
\end{equation}
where $\alpha = N b^2/6 > 0$ characterizes the chain connectivity, $\beta = -N^2 b^4 \tau/18 < 0$ represents the activity-induced stiffness for $\tau > 0$, and $\omega_\alpha(\br)$ is the self-consistent field for A and B components. Because \(\beta < 0\), the fourth-order term behaves as a \emph{negative diffusion} for high-frequency modes, which makes conventional explicit numerical schemes unstable. Importantly, this does not imply that the continuum equation is ill-posed; similar negative fourth-order terms appear in well-established models such as the Cahn-Hilliard equation \cite{Cahn1958}, where they describe physical phase separation but require implicit treatment for stability. To circumvent the instability here, we employ an implicit-explicit (IMEX) scheme to handle the linear operators.

We first adopt a splitting scheme that separates the linear and nonlinear operators \cite{Tzeremes2002}. Define
\begin{align}
    \mathcal{L} &\equiv \alpha \nabla^2 + \beta \nabla^4, \\
    \mathcal{N} &\equiv -\omega(\rvec).
\end{align}
The exact solution over a small time step \(\Delta s\) is approximated by
\begin{equation}
    q(s + \Delta s) \approx e^{\Delta s (\mathcal{L} + \mathcal{N})} q(s) 
    \approx e^{\frac{\Delta s}{2} \mathcal{N}} e^{\Delta s \mathcal{L}} e^{\frac{\Delta s}{2} \mathcal{N}} q(s),
\end{equation}
which is second-order accurate in time.

The nonlinear part \(e^{\frac{\Delta s}{2} \mathcal{N}}\) is a multiplicative operator that can be evaluated exactly in real space:
\begin{equation}
    e^{\frac{\Delta s}{2} \mathcal{N}} q(\rvec) = \exp\left[-\frac{\Delta s}{2} \omega(\rvec)\right] q(\rvec).
\end{equation}

An implicit-explicit (IMEX) scheme is employed to handle the linear operators:
\begin{itemize}
    \item Treat the second-order term (\(\alpha \nabla^2\)) explicitly (computationally inexpensive),
    \item Treat the fourth-order term (\(\beta \nabla^4\)) implicitly (ensures stability).
\end{itemize}

For the linear step \(q_{\text{lin}}(s + \Delta s) = e^{\Delta s \mathcal{L}} q_{\text{lin}}(s)\), we discretize as
\begin{equation}
    \frac{q^{n+1} - q^n}{\Delta s} = \alpha \nabla^2 q^n + \beta \nabla^4 q^{n+1},
\end{equation}
which yields the semi-discrete equation
\begin{equation}
    (1 - \Delta s \beta \nabla^4) q^{n+1} = (1 + \Delta s \alpha \nabla^2) q^n.
\end{equation}

Applying the Fourier transform \(\mathcal{F}\{q(\rvec)\} = \hat{q}(\kvec)\) and using the properties
\begin{align}
    \mathcal{F}\{\nabla^2 q\} &= -k^2 \hat{q}(\kvec), \quad k^2 = |\kvec|^2, \\
    \mathcal{F}\{\nabla^4 q\} &= k^4 \hat{q}(\kvec), \quad k^4 = (k^2)^2,
\end{align}
we obtain in Fourier space
\begin{equation}
    (1 - \Delta s \beta k^4) \hat{q}^{n+1}(\kvec) = (1 - \Delta s \alpha k^2) \hat{q}^n(\kvec).
\end{equation}
Note the minus sign in the right-hand side arises from \(\nabla^2 \rightarrow -k^2\).

Solving for \(\hat{q}^{n+1}(\kvec)\) gives
\begin{equation}
    \hat{q}^{n+1}(\kvec) = \frac{1 - \Delta s \alpha k^2}{1 - \Delta s \beta k^4} \hat{q}^n(\kvec).
\end{equation}
Since \(\beta < 0\), the denominator \(1 - \Delta s \beta k^4 = 1 + |\beta| \Delta s k^4 > 1\), guaranteeing unconditional stability for the linear operators.

Combining the operator splitting with the IMEX scheme, we obtain the following three-step algorithm:
\begin{itemize}
    \item \textbf{Step 1}: Multiplying the non-linear half operator on $q^n(\rvec)$ in real space:
    \begin{equation}
    q_1(\rvec) = \exp\left[-\frac{\Delta s}{2} \omega(\rvec)\right] q^n(\rvec).
\end{equation}
    \item \textbf{Step 2}: Applying the IMEX scheme for the linear part in Fourier space:
    \begin{align}
    \hat{q}_1(\kvec) &= \mathcal{F}\{q_1(\rvec)\}, \\
    \hat{q}_2(\kvec) &= \frac{1 - \Delta s \alpha k^2}{1 - \Delta s \beta k^4} \hat{q}_1(\kvec), \\
    q_2(\rvec) &= \mathcal{F}^{-1}\{\hat{q}_2(\kvec)\}.
\end{align}
    \item \textbf{Step 3}: Multiplying the rest non-linear half operator on $q_2(\rvec)$ in real space to obtain the $q^{n+1}(\rvec)$:
\begin{equation}
    q^{n+1}(\rvec) = \exp\left[-\frac{\Delta s}{2} \omega(\rvec)\right] q_2(\rvec).
\end{equation}
\end{itemize}

This numerical scheme successfully solves the modified diffusion equations while preserving computational tractability, enabling the exploration of steady-state morphologies in active polymer systems.

\subsection{Activity suppresses phase separation}

\begin{figure}
\includegraphics[width=1\columnwidth]{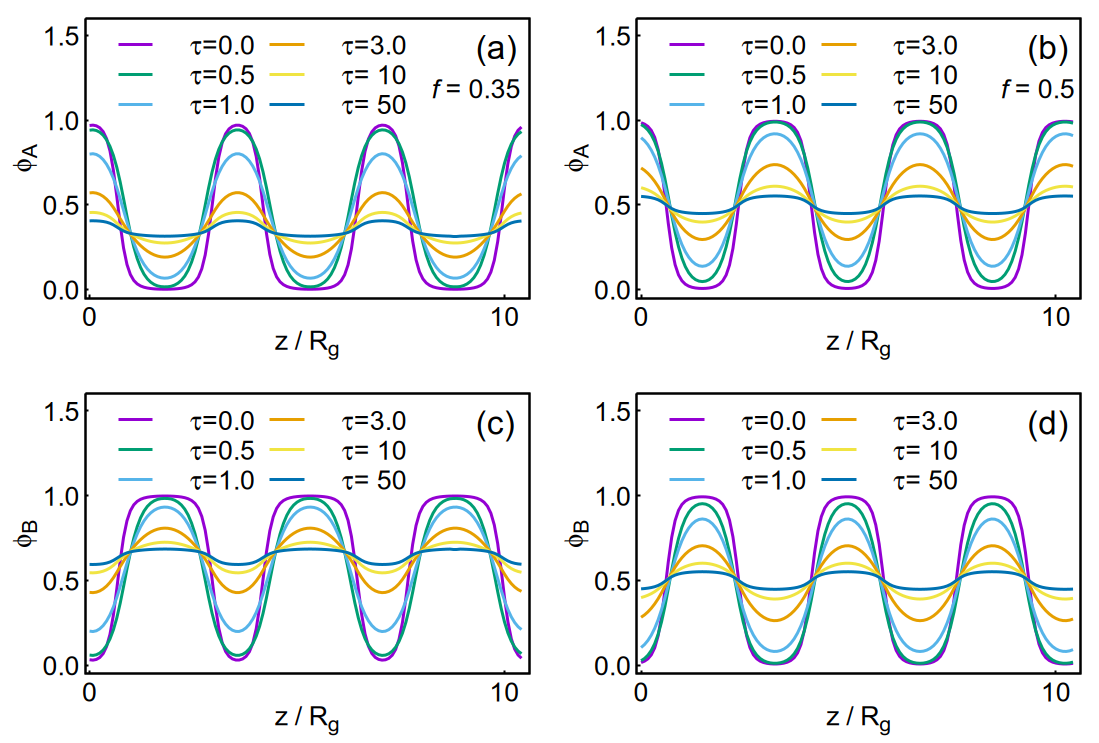}
  \caption{(a) Steady-state density profiles of the A component for an asymmetric block copolymer ($f=0.35$) with $\chi N=30.0$ in a one-dimensional simulation box of length $10.5 R_{g}$ at increasing activity $\tau = 0.0, 0.5, 1.0, 3.0, 10$, and $50$. (b) The same steady-state density profiles as in (a) but for a symmetric composition $f=0.5$. (c) Corresponding density profiles of the B component under the same conditions in (a).  (d) Corresponding B-component profiles for $f=0.5$.}\label{Fig1}
\end{figure}

We solve the one-dimensional active self-consistent field equations in a simulation box of length \(z = 10.5 R_g\), discretized with \(256\) grid points; the contour variable \(s\) is discretized into \(200\) steps. Here \(R_g\) is the equilibrium radius of gyration, which we take as the length unit \(l_0\). The Rouse relaxation time \(\tau_R\) serves as the time unit, so that the persistence time \(\tau = \tau_{\text{physical}} / \tau_R\) becomes dimensionless and quantifies the activity. We use the values \(\tau = 0,\;0.5,\;1.0,\;3.0,\;10,\;50\), ranging from the passive limit to strongly active conditions. The Flory--Huggins parameter is fixed at \(\chi N = 30.0\), and the fraction of A monomers is varied between \(f = 0.35\) and \(f = 0.5\).

Figure~\ref{Fig1}(a) and (b) display the steady-state density profiles of the A component for $f=0.35$ and $f=0.5$, respectively, at several values of $\tau$. For the passive case ($\tau=0$), the system exhibits a pronounced microphase-separated profile with well-defined A-rich and B-rich domains. As $\tau$ increases, the amplitude of the density modulation decreases systematically. For example, at $\tau=0.5$ the peak density of the A component is already visibly lower than in the passive case, and by $\tau=50$ the profile becomes almost flat. This trend indicates that activity suppresses the strength of phase separation. The same behavior is observed for the B component [Fig.~\ref{Fig1}(c) and (d)].

To quantify the suppression of phase separation, we plot the density difference $\Delta\phi_A = \phi_{A,\text{max}} - \phi_{A,\text{min}}$ as a function of $\tau$ for several block fractions (Fig.~\ref{Fig2}). For all compositions, $\Delta\phi_A$ decreases monotonically with $\tau$. At large $\tau$ ($\tau \gtrsim 1$), the decay follows a power law $\Delta\phi_A \propto \tau^{-1/2}$, as indicated by the dashed red line. This scaling is independent of the block fraction $f$, confirming that the activity-induced suppression of phase separation is a generic feature of the active Ornstein--Uhlenbeck polymer model. 

The results clearly demonstrate that activity, characterized by the persistence time $\tau$, tends to homogenize the system and reduce the amplitude of density modifications. This effect can be understood as arising from the additional non-equilibrium stresses that oppose the thermodynamic driving force for phase separation. Consequently, the active SCFT framework developed here provides a quantitative tool to predict how activity can be used to tune the morphology of block copolymer systems.

It is worth comparing our findings with the well-known motility-induced phase separation (MIPS) observed in dilute active particle suspensions \cite{Cates2015}. In such systems, activity generates clustering through the interplay between self-propulsion and steric repulsion. In contrast, our system is a dense incompressible diblock-copolymer melt, where the ordering is driven by Flory-Huggins interactions. Here, activity suppresses the pre-existing compositional ordering rather than inducing a new phase separation mechanism.

\subsection{Weak phase separation at large $\tau$}

\begin{figure}
\includegraphics[width=1\columnwidth]{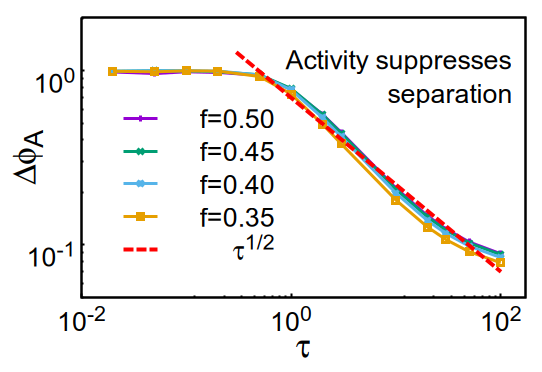}
  \caption{Density difference $\Delta\phi_A = \phi_{A,\text{max}} - \phi_{A,\text{min}}$ as a function of activity $\tau$ for several A-block fractions: $f = 0.35$, $0.40$, $0.45$, and $0.50$. The dashed red line indicates the theoretical scaling $\Delta\phi_A \propto \tau^{-1/2}$, demonstrating that increasing activity suppresses the amplitude of phase separation.}\label{Fig2}
\end{figure}

The numerical results presented in the previous subsection demonstrate that increasing activity (larger $\tau$) suppresses the amplitude of density modulations in microphase-separated active polymers. To gain further insight into the underlying mechanism, we perform an asymptotic analysis of the effective free energy functional of active OU polymers in the limit of large $\tau$. This analysis yields a scaling relation between the density difference and activity, which can be directly compared with the numerical results shown in Fig.~\ref{Fig2}.

For active Ornstein-Uhlenbeck polymers, the effective free energy functional is given by:
\begin{equation}
    F[\psi_A,\psi_B] = F_{\text{passive}} + F_{\text{active}} .
\end{equation}
The passive part is:
\begin{equation}
\begin{split}
    F_{\text{passive}} = -\ln (\frac{Q}{V}) + &\frac{1}{V}\int d\mathbf{r} \left[  \chi N \psi_A\psi_B - \omega_A\psi_A \right. \\
    & \left.- \omega_B\psi_B- \xi(1-\psi_A-\psi_B) \right],
    \end{split}
\end{equation}
and the active contribution is:
\begin{equation}
\begin{split}
    F_{\text{active}} = \frac{1}{V}\int d\br &\left[ \frac{\tau}{2N} |\chi N\nabla(\psi_A\psi_B)|^2 \right.\\
    &\left.- D\tau\chi N^2 \nabla^2(\psi_A\psi_B) \right].
    \end{split}
\end{equation}

For large $\tau$, we assume the system is in a weak phase separation regime.  Let
\begin{equation}
    \psi_A(\mathbf{r}) = f + \delta\psi(\mathbf{r}), \quad \psi_B(\mathbf{r}) = 1-f - \delta\psi(\mathbf{r}),
\end{equation}
with the incompressibility condition $\psi_A+\psi_B=1$ implying $\delta\psi_A=\delta\psi$, $\delta\psi_B=-\delta\psi$.  For the symmetric case $f=0.5$,
\begin{equation}
    \psi_A\psi_B = \frac{1}{4} - (\delta\psi)^2 .
\end{equation}
Consequently,
\begin{align}
    \nabla(\psi_A\psi_B) &= -2\delta\psi\nabla\delta\psi, \\
    |\nabla(\psi_A\psi_B)|^2 &= 4(\delta\psi)^2|\nabla\delta\psi|^2, \\
    \nabla^2(\psi_A\psi_B) &= -2|\nabla\delta\psi|^2 - 2\delta\psi\nabla^2\delta\psi .
\end{align}

Expanding the total free energy to fourth order (ignoring boundary terms) gives:
\begin{equation}
\begin{split}
    F = \frac{1}{V}\int d\br &\left[ A(\delta\psi)^2 + (B - 2D\tau\chi N^2)|\nabla\delta\psi|^2 \right.\\
    &\left.+ C(\delta\psi)^4 + 2\tau\chi^2 N (\delta\psi)^2|\nabla\delta\psi|^2 \right],
    \end{split}
\end{equation}
where 
\begin{equation}
    A = -\chi N + \frac{1}{2f(1-f)N}
\end{equation}
is obtained from RPA analysis \cite{Daoud1975}, $B$ is related to chain elasticity, and $C$ is the fourth-order coefficient.

Assume a single-mode modulation:
\begin{equation}
    \delta\psi(\mathbf{r}) = \epsilon \cos(qz),
\end{equation}
where $\epsilon$ is the amplitude and $q$ is the wave number.

Spatial averages are:
\begin{align}
    \langle (\delta\psi)^2 \rangle &= \frac{\epsilon^2}{2}, \\
    \langle |\nabla\delta\psi|^2 \rangle &= \frac{\epsilon^2 q^2}{2}, \\
    \langle (\delta\psi)^4 \rangle &= \frac{3\epsilon^4}{8}, \\
    \langle (\delta\psi)^2|\nabla\delta\psi|^2 \rangle &= \frac{\epsilon^4 q^2}{8}.
\end{align}

The average free energy density becomes:
\begin{equation}
    \bar{F} = \frac{A}{2}\epsilon^2 + \frac{B}{2}\epsilon^2 q^2 + \frac{3C}{8}\epsilon^4 + \frac{\tau\chi^2 N}{4} \epsilon^4 q^2,
\end{equation}
where $B = B_0 - 2D\tau\chi N^2$.  First by minimizing with respect to $q$, we calculate:
\begin{equation}
       q^2 = -\frac{2B}{\tau\chi^2 N \epsilon^2}.
\end{equation}
For large $\tau$, $B<0$ so $q^2>0$.  Insert $q^2$ into $\bar{f}$:
\begin{equation}
    \bar{F} = \frac{A}{2}\epsilon^2 - \frac{B^2}{\tau\chi^2 N} + \left( \frac{3C}{8} - \frac{B}{2} \right)\epsilon^4 .
\end{equation}
Now minimizing with respect to $\epsilon$, we get:
\begin{equation}
   \epsilon^2 = -\frac{A}{\frac{3C}{2} - 2B_0+ 4D\tau\chi N^2}.
\end{equation}
When $\tau$ is large, the term $4D\tau\chi N^2$ dominates the denominator, giving
\begin{equation}
    \epsilon^2 \sim \frac{1}{\tau}
    \quad \Rightarrow \quad
    \epsilon \sim \tau^{-1/2}.
\end{equation}
Thus the density difference scales as
\begin{equation}
    \Delta\phi = 2\epsilon \propto \tau^{-1/2} .
\end{equation}

This theoretical scaling relation, $\Delta\phi \propto \tau^{-1/2}$, is in excellent agreement with the numerical results from the active SCFT calculations. As shown in Fig.~\ref{Fig2}, for $\tau \gtrsim 1$, the density difference $\Delta\phi_A$ indeed follows a power-law decay with an exponent of $-1/2$, as indicated by the dashed red line. The theoretical curve (red dashed line) closely matches the numerical data points across all studied block fractions $f$, confirming the validity of the asymptotic analysis.

From the expression for $\epsilon^2$,
\begin{equation}
    \epsilon^2 \approx -\frac{A}{4D\tau\chi N^2},
\end{equation}
where $A = -\chi N + \frac{1}{2f(1-f)N}$.  At large $\tau$, the denominator $4D\tau\chi N^2$ dominates, suppressing the $f$-dependence of $A$.  Consequently, the decay exponent $-1/2$ is universal for all $f$, while the prefactor varies only slightly with $f$.  This theoretical prediction explains the numerical observation that for $\tau \gtrsim 1$, the decay of $\Delta\phi$ becomes nearly independent of $f$, as seen in Fig.~\ref{Fig2} where the curves for different $f$ collapse onto a single scaling line at large $\tau$.

In summary, the asymptotic analysis of the free energy functional in the large-$\tau$ limit provides a clear theoretical foundation for the observed suppression of phase separation. The predicted scaling $\Delta\phi \propto \tau^{-1/2}$ is robust and independent of the block copolymer composition, highlighting the dominant role of activity in homogenizing the system.

\section{Conclusion}\label{Conclusion}

In this work, we have developed an active self-consistent field theory (ASCFT) that extends the classical equilibrium framework to non-equilibrium steady states of active Ornstein-Uhlenbeck polymers. The theory is founded on the steady-state probability distribution derived under the Unified Colored Noise Approximation, from which we construct an effective Hamiltonian incorporating both passive interactions and activity-driven contributions. The resulting free energy functional explicitly couples the persistence time $\tau$ of the colored noise with the Flory-Huggins parameter $\chi N$, providing a unified mean-field description for active polymers.

Numerical solution of the modified diffusion equations -- enabled by a stable implicit-explicit operator splitting scheme -- reveals that activity profoundly suppresses micro-phase separation. For symmetric and asymmetric block copolymers, increasing $\tau$ systematically reduces the amplitude of density modulations, leading to system homogenization. The density difference follows a power-law decay $\Delta\phi_A \propto \tau^{-1/2}$ at large $\tau$, a scaling behavior that is independent of block composition. Asymptotic analysis of the free energy functional confirms this scaling and explains its universality, demonstrating close agreement between theory and simulation.

The ASCFT framework establishes that kinetic activity, parameterized by $\tau$, can effectively compete with and counteract thermodynamic driving forces for phase separation. This offers a novel mechanism to control polymer self-assembly via non-equilibrium driving. By bridging polymer field theory with active matter physics, our work provides a foundation for exploring more complex architectures, hydrodynamic couplings, and time-dependent activity patterns in synthetic and biological soft materials.

\begin{acknowledgments}
The authors acknowledges the stimulating discussions from Prof. Marcus M\"uller, and the passive SCFT code from Prof. Weihua Li. The financial support from the National Natural Science Foundation of China under Grant Nos. 12374207, 12347102, 12174184, the Fundamental and Interdisciplinary Disciplines Breakthrough Plan of the Ministry of Education of China (JYB2025XDXM502), the Natural Science Foundation of Jiangsu Province (BK20233001), and the Innovation Program for Quantum Science and Technology (2024ZD0300101) are acknowledged. We also thanks the super computing resources at the Big Data Computing Center of Southeast University and the Beijing Super Cloud Computing Center (BSCC).
\end{acknowledgments}

\bibliography{bibtex}

\end{document}